# A practical multi-party computation algorithm for a secure distributed online voting system

Juanjo Bermúdez
Non-affiliated
c/Emigrant 30 Bajo 2
L'Hospitalet. BCN, Spain 08906
juanjo75es@gmail.com

## ABSTRACT
We present an online voting architecture based on partitioning the election in small clusters of voters and using a new Multi-party Computation algorithm for obtaining voting results from the clusters. This new algorithm has some practical advantages over other previously known algorithms and isn't bound to any specific cryptographic concept; so it can be adapted to future cryptographic exigencies. Compared with other online voting technologies, we see that this new architecture is less vulnerable to hacker attacks and attacks from dishonest authorities, given that no sensitive information is stored in any public server and there is no need for any trustee to safeguard the legality of the election process. Even in case of an attack succeeding, the risks associated with the overall election are far lower than with any other voting system. This architecture can also be combined with any other voting system, inheriting advantages from both systems.

## 1. INTRODUCTION
Centralized online voting systems are commonly considered to be insecure. Some centralized online elections have taken place in which different vulnerabilities have been later detected. As example, Clash Vulnerability [3] was detected for the prominent Helios voting system [5] after it was used by Université Catholique de Louvain and the International Association for Cryptologic Research. Many experts think that we are far beyond the day that these technologies will be ready for massive deployment, if ever they are.

An alternative method for online voting is making use of a distributed architecture instead of a centralized one. These methods have not been deployed on real use cases because of their practical limitations, despite being largely studied in scientific literature and some studies have even proven their feasibility for this task [1, 4, 6, 7]. Some practical limitations are related with the high number of messages that these methods need to exchange between nodes. This number usually grows exponentially with the number of voters.

We propose a voting architecture based on: (i) dividing the election in small clusters of voters that participate in independent partial elections, where results are later joined to get the global election result; (ii) making use of a new Multi-party Computation (MPC) algorithm with no practical limitations for this scenario.

## 2. IDEAL PROPERTIES FOR AN MPC VOTING ALGORITHM
For a Multi-party Computation Voting Algorithm to be practical for such scenario (dividing the election in small clusters of voters), some properties need to be satisfied.

**Not needing trusted authorities**. If there is an authority supervising every voting taking place, even if it's a different authority for every voting, we are only dividing the problem but it's still the same problem. It makes it harder for an attacker to have a big impact with his attack, but it also makes it a lot more complex to organize the voting and you are creating a lot more (smaller) points of failure: an attack to some of these points could have a small impact but still alter the outcome on tight elections. Hence, in an ideal scenario, there would be no authority supervising the cluster's votings. The votings need to rely completely on the properties of the algorithm, without any privileged node in which other nodes need to deposit trust. This requisite discards some MPC algorithms based on homomorphic operations [2].

**Safeguarding vote integrity**. No one should be able to modify someone else's vote or break the rules of the voting even without trustees supervising the process.

**Safeguarding vote confidentiality**. If this requirement is not in place, it is trivial to build a voting architecture satisfying all other requirements. Some MPC algorithms rely on a teller node which receives encrypted shares from every voter node [2]. These systems are usually susceptible to attacks from this teller node consisting of: (i) taking a number n of votes and computing the result, (ii) adding a new vote to the same group of voters, and (iii) computing a new result that proves the option voted by the last added voter. If you repeat this process, you can find out every single vote processed by the teller.

**Being able to identify dishonest voters trying to alter the result or violate confidentiality**. Any attempt to corrupt the integrity of an election must be detected and correctly attributed. It's not only needed that the algorithm detects and avoids attempts to violate integrity or confidentiality of the voting, but also that it identifies the users who are trying to abuse. This way you can take disciplinary actions against these users for: (i) avoiding DoS attacks consisting of an infinite loop of voting cancellations, (ii) diminishing the incentives for attacking the election, by increasing the risk of punishment. Without that requirement, the system could be unreliable.

**Requiring a reasonable processing time and requirement of resources, even when dishonest voters are present**. Without this requirement, the process would not be practical; many voters would reject using it because of long waiting times or it would be too expensive. To satisfy this requirement every step in the algorithm should take a small portion of time: it would be desirable that in no case the cluster is blocked too long waiting for a node to make a calculation. Some algorithms satisfy all conditions to be a valid algorithm for this voting architecture except this one. Not satisfying this condition would imply that at any moment a node could take a long time to make a calculation, only to disconnect before returning a result. It would make the voting to be cancelled after a long wait and no one could distinguish intended disconnections from casual ones. If repeated by a large number of users, it could block an election making it impossible to vote, to many (or most) users. We have to think that the timeout needs to enable even the participation of a node with low computational power and low bandwidth.

# 3. PROPOSED SYSTEM

We propose a new MPC algorithm that satisfies these requirements. It doesn't make use of any new cryptographic procedure or theory; any well-known (present or future) and tested cryptographic procedure to establish a secure channel between nodes can be used. It also doesn't depend on trust on any privileged node. All nodes of the cluster cooperate to calculate the result, making the same kind of calculations on every node.

The algorithm is decomposed in two stages.

## 3.1 Stage 1

1. A list of virtual ballots (v-ballots) is created. Every v-ballot has its own id. There are sc*(k+1) different v-ballots for every option available to vote, being sc the number of nodes (voters) connected to the cluster, and k a constant that typically will be 1. Hence, if there are 2 available options, with 10 voters connected, and we fix k as 1, there will be 20 v-ballots for option 1 and 20 v-ballots for option 2. We can optionally add 20 v-ballots for the abstention option. We will call ao the total number of available options (including abstention).

2. There are ao*k+1 rounds where every node extracts a v-ballot from the list and passes the list of remaining v-ballots to the next node. Usually there will be a pre-established order determined before the start, and every node will check that the established order is not broken. Every node will also check that the list of remaining v-ballots received on every round doesn't have any inconsistency with previous rounds (i.e.: a v-ballot that wasn't in the list in a previous round isn't now in the list). If any inconsistency is detected, the voting is cancelled and a warning is assigned to the node reporting the inconsistency (as he could be lying) and to the two possible cheating nodes determined by another algorithm. For clarity and simplicity, we consider this algorithm out of the scope of this paper. RULE A: Every node must extract a v-ballot for every available option and an additional v-ballot for the option they want to vote for. So at the end of the ao*k+1 rounds, every node must have a v-ballot for every option plus a v-ballot for the option chosen.

3. The list of the remaining v-ballots is made public by the last node. Every node checks that this list is consistent with his data. This list will be used to determine the final result.

It could be that any node had cheated breaking RULE A; for example, selecting three v-ballots for the option they want as winner and none for another option. Hence, stage 2 consists of detecting any node which could have broken this rule.

## 3.2 Stage 2

There are some possible implementations of this stage 2. Here we will describe the one we usually use for our tests.

1. Every node asks every other node (or an established high number of them), through a secure channel, for the id of a v-ballot they have selected in stage 1. The asker node imposes the option the returned v-ballot needs to pertain to. These ids, if the nodes follow the protocol, are only known by the nodes receiving them. Note that knowing only one v-ballot from every node doesn't allow you to know what option he selected, as you know for sure that if he didn't cheat, he selected at least one v-ballot for every option.

2. For every v-ballot received, every node has to check:
   A) that the v-ballot is not one of the v-ballots in the list of remaining v-ballots made public after the election;
   B) that the v-ballot is not one of the v-ballots that it (the node asking) secretly selected in stage 1;
   C) that the v-ballot is not one of the v-ballots that other nodes have reported to have selected.

3. If no collision of v-ballots is reported, the nodes agree to declare valid the voting and sign the result. The result is calculated this way: the number of votes for every option is n minus the number of v-ballots in the list of remaining v-ballots corresponding to the option. If a collision is reported, the nodes being in doubt, including the one reporting the collision, are annotated with a warning.

Table 1 illustrates the algorithm for an over-simplified case (CASE 1) with 4 voters (A,B,C,D), with 2 options available to vote (R, N), and k=1.

Table 1

|  | Voter A | Voter B | Voter C | Voter D | Remaining v-ballots |
|---|---|---|---|---|---|
| Round 1 | N2 | N1 | N6 | R3 | N3,N4,N5,N7,N8 R1,R2,R4,R5,R6,R7,R8 |
| Round 2 | R1 | R8 | R4 | N5 | N3,N4,N7,N8 R2,R5,R6,R7 |
| Round 3 | N8 | N3 | R2 | R7 | N4,N7 R5,R6 |

# 4. RISK ANALYSIS

We will now calculate the risks associated to the algorithm and compare them with the risks associated to a centralized online voting and a traditional paper ballot system.

First, we will calculate the risk that a dishonest node succeeds with altering the voting integrity, making the rest of the nodes of the cluster to accept a fraudulent result. For CASE 1, there are two ways in that a node could try to alter the result: (i) CHEAT 1: he could always select a v-ballot from the same option, or (ii) CHEAT 2: he could modify the list of remaining v-ballots that he receives on any round, as if previous voters would have selected other ones.

## 4.1 Risk of cheat 1 with a single attacker

Hence, supposing we are in CASE 1, and node A is the one trying to make CHEAT 1, instead of choosing N2,R1,N8 he would choose for example, N2,N4,N8. So the final remaining v-ballots in CASE 1 would be: N7,R1,R5,R6. This result would make us think that 3 voters voted for N, when really only nodes A, and B voted for N.

Now Node A would have to pass stage 2, where he will be asked one time for an N v-ballot and two times for an R v-ballot (could also be one R and two N but let's suppose it this way this time). When he is asked for an N v-ballot he will simply return a randomly selected one from the three N v-ballots he selected. If there aren't more nodes cheating, it will pass the test as no node will have the same v-ballot. When asked for an R v-ballot he needs to lie given that he didn't extract any R v-ballot. If he is lucky, the node asking him will not discover that he is lying. If he is not, the node asking him will cancel the voting and initiate a new algorithm to detect the two nodes in conflict so that the nodes causing the conflict get a warning (the reporting node will also get a warning as he could be lying). This algorithm will not be explained in this paper for simplification.

In this case (CASE 1), the probability that Node A passes the test having lied can be calculated.

$$p\_collision = p\_collision\_with\_asker + p\_collision\_with\_another\_node$$

The probability of collision detected in every single check can be calculated this way:

If the node asking selected it, it's 1.

If the node that selected it also voted for this option, it's the probability that he returned it when asked: ½.

If the node that selected it didn't select it as his vote, it's 1, because it's the only value that it can return.

Hence, we can simplify it as p_collision≥0,5 and p_no_collision ≤0,5.

Given that there will be another node asking, the probability of no collision on first check and no collision on second check is lower (≤) than 0,5*0,5, so it is ≤0,25.

So a dishonest node in this case would have less than a 25% chance of altering the result without being detected.

A 25% chance is still too high, but we have to remember that it's an over-simplified case. We can learn from this data that we should not make clusters with only 4 nodes.

The general formula for the minimum probability of a node to cheat and not be discovered, p_cheat is:

$$p\_cheat(p\_no\_collision, ao, sc) = p\_no\_collision^{sc/ao}$$

Table 2 shows results for different parameters.

Table 2

| sc | ao | p_cheat |
|---|---|---|
| 4 | 2 | 0,25 |
| 8 | 2 | 0,06 |
| 15 | 2 | 0,006 |
| 25 | 2 | 1,7E-4 |
| 40 | 2 | 9,5E-7 |
| 4 | 3 | 0,40 |
| 8 | 3 | 0,15 |
| 15 | 3 | 0,03 |
| 25 | 3 | 0,003 |
| 40 | 3 | 9,7E-5 |
| 4 | 5 | 0,57 |
| 8 | 5 | 0,33 |
| 15 | 5 | 0,12 |
| 25 | 5 | 0,03 |
| 40 | 5 | 0,004 |

So, for example, if you organize an election using this system, establishing 25 nodes per cluster, and having 3 available options to vote (including abstention), a cheater knows he has a greater than 99,7% probability of being discovered. Supposing that after 3 warnings a punishment is established, he would have a greater than $0,997^3$ probability of being punished. That's a greater than 99,1% probability of being punished, with a possible reward consisting on altering one single vote.

We also have to recall that we established in stage 2 that every node would ask one single option to every node. It's a very conservative requirement. We could make every node ask the other nodes for a v-ballot for every option and they would still not be able to determine what options the other nodes voted for. In this case, the probability for a cheater to pass the checks would always be less or equal than if there were only 2 options to vote. Therefore, in this case, p_cheat for 25 nodes and 3 options would be under 1,7E-4 and he would have an over 99,95% probability of being punished.

## 4.2 Risk of cheat 1 with multiple attackers

We have studied a case with only one dishonest node trying to alter the result. In case more than one node is dishonest in this way, there are two options:

- They are independent attackers that will report any cheat from other nodes
- They are attackers acting in coordination and they won't report collisions from other attackers.

In case they are independent attackers, the individual chances of success for every attacker are the same as in the studied case. But the probability that none of the dishonest nodes is detected it's still lower, as the probability of being discovered is the sum of the probabilities for every cheating node. Being dn: number of dishonest nodes, the probability of cheating success is $p\_cheat^{dn}$. So if p_cheat is 0,003 and there are 3 attackers, it's now 2,7E-8.

In case they are coordinated attackers, the new probability is

$$p(c) = p(no\ collision)^{(n-dn)/p}$$

Hence, in case k=1, p=25, o=3, where p_cheat was under 0,003; it's now under 0,006 for every dishonest node if nd=3. But still they have to succeed all three nodes, so the new probability is under $0,006^3=2,1E-7$. If nd=20, the new p_cheat is 0,31 and the probability that none of the 20 dishonest nodes is discovered is under $0,31^{20}=6,7E-11$.

## 4.3 Risk of cheat 1 with multiple attackers but only one active cheater

There is also the possibility that only one of the dishonest nodes cheats and the rest of dishonest nodes only provide information to this node. They wouldn't report collisions when a dishonest cooperator is involved. In that case, for k=1,p=25,o=3,nd=20, p_cheat would be under 0,31, and given that all other dishonest nodes will not cheat, the total probability of success will be under 0,31. So if 20 dishonest coordinated nodes attack a cluster, they could alter one vote with an under 31% probability. The attacker would still have a greater than 69% probability to be discovered. If the same parameters are repeated for 3 consecutive votings, and 3 warnings suppose a punishment, the dishonest node will have an over 32% chance of being punished, and therefore, an under 68% chance of success. As we will see next, even with this quite high success ratio, it's very unlikely that any online voting will derive to such scenario where there are 20 coordinated cheaters for every 25 voters. Even if this happens, two counter-measures will be exposed.

### 4.3.1 Risk of attackers' concentration

Let's see next what circumstances are needed so that an online voting derives to a situation where 20 out of 25 voters are coordinated cheaters. It could be that 20 out of 25 of the total number of voters were cheating coordinately and it could be that at a certain moment there were a high number of coordinated cheaters connected at the same time. Hence, supposing it's an election with 22 million voters, a voting period of 12 hours, and an average of 4 minutes to vote, it would mean that at any moment there would be around 122,000 voters voting. To achieve a percentage of 20 out of 22 voters cheating coordinately, 500,000 cheaters would need to be connected at the same minute coordinately. They would successfully alter around 16,924 votes, and around 7,964 attackers would be punished. Even if this unlikely scenario takes place, a counter-measure could simply consist of rewarding anyone reporting the attack. Attackers will need a way to coordinate their attack, and if only one person informs the authorities on the attack, all attackers could be identified and punished. In massive online elections with high interests in play, authorities would need to make an effort to identify these kind of attacks. In that case, the risk for attackers would be so high, and the reward so low (altering a small percentage of votes) that it would be a very unlikely attack. For small voting, where authorities can't afford the expenses for detecting such coordinated attacks, and where it's more likely that in some cases a high percentage of voters pretend to cheat, a counter-measure would be to vote in parallel with a centralized online voting method. At the end of the election you will have the result obtained by distributed voting and the result obtained by centralized voting. You are increasing the chances of being attacked under the centralized architecture, but this way you know that an attack to the central server or a coordinated attack by a massive number of voters in the distributed architecture will be in any case detected, the result will be discarded, and attackers will not get any reward.

## 4.4 Risk of cheat 2

Next, we will study probabilities for attacks of type CHEAT 2. Supposing we are in CASE 1 and node C is the cheater, he is receiving at the second round these v-ballots: N3,N4,N5,N7,N8,R2,R4,R5,R6,R7. The cheat would consist, for example, on replacing one of the v-ballots from one option for a v-ballot from another option. He could, in this example, select R4, and replace R5 for N2, making it seem like if the node who selected N2 selected R5 instead. Now node D gets the list of v-ballots, he doesn't detect any inconsistency, and he selects N5. Now Node A receives N2,N3,N4,N7,N8,R2,R6,R7 and he notes that N2 is in the list despite him selecting it. So the cheat fails. The only way it could have worked at this stage would be if Node D had selected N2. Let's suppose it's the case. Node A would receive: N3,N4,N5,N7,N8,R2,R6,R7. There is no inconsistency, so he will continue. He selects N8. Then Node B selects N3, Node C selects R2, and Node D selects R7. The final remaining v-ballots are: N4,N5,N7,R6. Apparently 3 nodes would have voted for R and one for N. Cheat 2 would succeed at this stage in this case. But then in stage 2 the probabilities of not being discovered would be the same as that in the case where the cheater selects 2 v-ballots from same option. It's the same circumstance to be detected: two nodes selecting the same v-ballot. So, the same table applies and, in example, if we establish 25 nodes per cluster, and 3 available options to vote (including abstention), a cheater knows he has a probability greater than 99,7% of being discovered, and a probability higher than 99,1% for being punished if he tries again to cheat. The possible reward again would be altering only one vote. In case the cheater modifies more v-ballots from the list he receives, the probability of being discovered and punished would be still higher. Note that we didn't explain how we would know which nodes should be warned for being possible cheaters. A new iterative algorithm will be used to detect the nodes candidates for cheating. For avoiding unnecessary complication and extension of this paper, we won't explain this algorithm in that occasion. You can consider for simplicity that every node is warned and they aren't connected again on the same cluster to avoid the possibility that the cheater node sweeps along the other nodes to a punishment.

## 4.5 Risk of privacy violation

We will study next the risk of a secret vote being violated. Let's start supposing the voting has 3 available options; k is 1, cluster size is 15 nodes, and there are two malicious voters in the same cluster who want to cooperate to find out votes from other voters. One strategy to cooperate would be to share information from the verification stage of the algorithm. This way, they could find out that a node selected two v-ballots with different id's corresponding to the same option. It would mean that this node voted for this option. In order to make this strategy work, they should ask every node for v-ballots from the same option. The option that every node asks to other nodes is determined by their ordered position in the list of nodes. This order is randomly established before starting the vote, so there is a probability p_nosame that they don't ask for the same option. This probability p_nosame, if ao>=nt is

$$p\_nosame(no, nt) = \frac{(ao-1)!}{(ao-nt)!} \Big/ ao^{(nt-1)}$$

, and if ao<nt it's

$$p\_nosame(ao, nt) = 0$$

So, the probability that at least two nodes ask for the same option is p_same= 1 – p_nosame. In this case it would be 0,33

Here we see a table with different values:

Table 3

| #options (ao) | #cheaters (nt) | p_same |
|---|---|---|
| 3 | 2 | 0,33 |
| 3 | 3 | 0,78 |
| 5 | 2 | 0,2 |
| 5 | 4 | 0,81 |
| 5 | 6 | 1 |

Therefore, if we organize an election with 3 available options, when 2 cheaters are in the same cluster, they have a 33% chance of finding out some votes; if there are 3 cheaters it's a 78% chance, and if there are more than 3 cheaters it's a 100%.

Next we will calculate how many votes they could find out.

We will call p_diffids(nt) to the probability that any node asked for an v-ballot by nt cheaters who ask the same option, respond with 2 different v-ballots (proving he voted that option). The formula to calculate it is:

$$p\_diffids(nt) = 1 - 0.5^{(nt-1)}$$

We will call p_match(ao) to the probability that any node asked for a v-ballot corresponding to a specific option in stage 2, already voted for that option. We can't know this probability but we will suppose that on average it is 1/ao.

Hence, in a cluster where there are *nt* cheaters and *ao* options to vote, the probability for every node to reveal his vote to these *nt* cheaters would be:

$$p\_reveal(ao, nt) = p\_same(ao, nt) \times p\_diffids(nt) \times p\_match(ao)$$

Table 4

| ao | nt | p_reveal(ao,nt) |
|---|---|---|
| 3 | 2 | 0,06 |
| 3 | 3 | 0,19 |
| 5 | 2 | 0,02 |
| 5 | 4 | 0,14 |
| 5 | 6 | 0,19 |

Combining these two probabilities, we can make a table with the average number of votes expected to be discovered for different parameters.

$$\#discovered = \left(\#attackers - \left(\#attackers * \left(\frac{nt}{cs}\right)\right)\right) * \frac{p\_reveal}{2}$$

Table 5

| ao | nt | cs | #attackers | p_reveal | #discovered |
|---|---|---|---|---|---|
| 3 | 2 | 15 | 1000 | 0,06 | 24 |
| 3 | 3 | 15 | 1000 | 0,19 | 77 |
| 3 | 4 | 15 | 1000 | 0,29 | 107 |
| 5 | 2 | 35 | 1000 | 0,02 | 9 |
| 5 | 3 | 35 | 1000 | 0,08 | 35 |
| 5 | 4 | 35 | 1000 | 0,14 | 62 |
| 5 | 5 | 35 | 1000 | 0,18 | 77 |
| 5 | 6 | 35 | 1000 | 0,19 | 80 |

This means that if an organized team of 1000 voters attack an election with 3 available options (a*o*=3), and 15 nodes per cluster, they could find out around 107 votes, and in case *ao* was 5 and cs=35, around 80 votes. This is a concern for a real voting but there is a way to avoid this risk. We will see it next.

## 5. OVERALL ARCHITECTURE AND PROPERTIES OF THE VOTING SYSTEM

We have found out that there is a security risk concerning the privacy of a vote. There is also a risk that nodes exposing their IP to other nodes suffer different DoS attacks or attempts of intrusion. To solve these problems we propose a distributed architecture, where nodes operate in clusters independently, but where they connect through intermediary nodes provided by an authority. These nodes will supervise voter legitimacy to vote, shallow IPs of voters, record warnings to users, apply punishments, and finally, shallow user ids offering temporal random ids to them during the voting. User-id-shallowing makes irrelevant the possibility that a group of attackers violate privacy from some nodes, given that they will know their shallowed id but not the real id. These intermediary nodes will transmit every message between nodes but they will not have access to the contents of these messages: every message will be encrypted with a public key that only the receiver can decrypt. The nodes will verify that intermediary nodes behave as expected, invalidating the voting if they detect any improper operation.

Hence, the overall properties of the proposed voting architecture are:

- If appropriate parameters are applied, the risk of altering votes for sporadic attackers is near zero, with very high risk of being punished for the attackers.
- If appropriate counter-measures are implemented, even with a high percentage of voters being attackers, the probability that they succeed in the attack is near zero.
- Risk of breaking the privacy of votes is zero if attackers only operate coordinately on clusters.
- In case an attacker is able to get unauthorized access to the intermediary nodes, it will not alter any vote nor find out any vote. The only information stored by these nodes is the information that in the end will be made public to anyone who wants to download it.
- In case an attacker is able to get unauthorized access to the intermediary nodes and operating coordinately with a number of non-honest nodes in different clusters, he could break the privacy of a number of voters numbered, if appropriate parameters are selected, not over approximately one-tenth of the number of coordinated attackers. In that case, they would know what these nodes voted for but they would not have any way at all to demonstrate it; so a public leaking of data would be useless, as this data could be falsified.

- At the end of the election, the result of every cluster is published. Every result is signed by every voter. So, anybody can check that every voter signed only one result, and every voter can check that the result of his cluster is published. In case a result was not published, any voter in the cluster could resubmit the result. This result would be signed by every voter and the intermediary nodes that participated in the voting, proving that this is a valid result and has to be included in the counting. Anyone could also check that all signatures correspond to valid voters in the census.
- The number of messages exchanged between nodes can grow linearly with the number of nodes: O(N), depending on some configuration parameters. It's an advantage over other known MPC algorithms that usually have O(N^2). All messages between nodes should take no longer than 1 second to be processed and transmitted, so you can have a timeout to avoid long waits. Long waits would be a way to enable DoS attacks. So our voting system is also reliable.
- Dishonest users attempting to alter votes can be identified and punished; but we don't expose in this paper the algorithm we employ to do it efficiently.

There is still one kind of attack that could affect our voting system (and any other known online voting system). If a malware spreads through internet, being undetected, and including code to alter the voting client software, it, in theory, could make affected voters lose their privacy and the inviolability of their vote. There are many practical reasons why it's quite unlikely that this kind of attack would succeed as the malware would need to spread fast and massively and stay undetected (speed and stealth are two contradictory requirements). But in case that happens, there is still one advantage for our system that would make it possible to detect the attack and invalidate the result. Given that you don't have all votes randomly mixed, but only mixed in small sets, you could make a statistical analysis of the result, comparing a subset of the stored results with results obtained for these voters making use of an alternative voting system. If there is a statistically significant difference between the result obtained with our voting system and the partial results obtained with another system, anyone could appeal the result of the election. These votings with alternative systems could take place either after the election, or in parallel to the election.

Finally, we can compare these risks with the risks associated to other voting systems.

We know that centralized online voting systems often have high risks in relation to dishonest authorities and hacker attacks. Any of these attacks could usually cause the result to be totally altered and nobody ever knowing it. Also, every single vote could be disclosed.

We also know that common voting systems based on paper ballots require very rigid protocols to minimize the risk of manipulation, which anyway can't be totally granted for sure in any election. Observers usually consider a percentage of breaches in the protocol as acceptable; otherwise every election would need to be invalidated.

Table 6 is a visualization of a comparison of risks.

Table 6

|  | Paper ballots | Centralized online voting | Our distributed voting system |
|---|---|---|---|
| Dishonest users | Medium risk | Low risk | Low risk |
| Dishonest authorities | Medium risk | High risk | Low risk |
| Hacker attack | Low risk | High risk | Low risk |

- Low risk
- Medium risk
- High risk

## 6. CONCLUSIONS

We propose a voting system with security advantages over other known voting systems. The focus is on minimizing risks instead of minimizing the probability of attack success, so even if an attack succeeds, the risk for the overall election is lower than in any other voting method. Even then, the probability of an attack to succeed is in most cases, near zero. It can be used in an online election as the only method for voting, or used in conjunction with a well-known centralized online voting technology, getting two results using totally independent voting systems, with different risks associated. If used in conjunction, you inherit the risks related to privacy from both the systems but you drastically reduce the risk of an undetected fraudulent result. Despite inheriting risks from both systems, these risks are not a serious concern compared with the risks associated to using only a centralized voting system. So, we propose the convenience of using this new distributed voting concept in conjunction with any well-established online voting technology whenever making use of an online voting technology. This extra security could strongly facilitate the extension of the use of online voting systems.